\documentclass{article}

\usepackage[numbers]{natbib}
\usepackage{graphicx}
\usepackage{amsmath}
\usepackage{amssymb}
\usepackage{latexsym}

\usepackage[margin=1in]{geometry}

\usepackage{rotating}

\begin{document}
\bibliographystyle{aip}

\begin{center}
  {\Large High level modeling of tonic dopamine mechanisms in striatal neurons}\\
\vspace{1cm}
Mark D. Humphries\\

{Department of Psychology, University of Sheffield \\ Sheffield, S10 2TP UK}\\
m.d.humphries@shef.ac.uk \\

Technical Report: ABRG3, November, 2003 \\

\end{center}
\begin{abstract}
  The extant versions of many basal ganglia models
  use a `gating' model of dopamine function which enhances input
  to D1 receptor units and attenuates input to D2 receptor units.
  There is evidence that this model is unsatisfactory because (a)
  there are not sufficient dopaminergic synapses to gate all input and
  (b) dopamine's main effect is likely to be on the ion-channels
  contributing to the neuron's membrane potential. Thus, an
  alternative output function-based model of dopamine's effect is
  proposed which accounts for the dopamine-mediated changes in
  ion-channel based currents. Simulation results show that the
  selection and switching properties of the intrinsic and extended
  models are retained with the new models. The parameter regimes under
  which this occurs leads us to predict that an L-type Ca$^{2+}$
  current is likely to be the major determinant of striatal neuron
  output if the basal ganglia is indeed an action selection mechanism.
  In addition, the results provide evidence that increasing dopamine
  can improve a neuron's signal-to-noise ratio.

\end{abstract}


\section{Introduction}
The action of dopamine at its receptor sites in the striatum has been
a source of continuing controversy since it was first established that
striatal cells received direct catecholaminergic input from the
substantia nigra pars compacta \citep{POIRIER:1965}. Primarily the
controversy has centered around how dopamine affects striatal output
through binding at the two main dopamine receptor sub-types, D1 and
D2, expressed by striatal spiny (or projection) neurons.

The prevailing viewpoint is that dopamine acts to attenuate spiny
neuron output via D2 receptors and enhance it via D1 receptors and
that these receptor types are found on two distinct subpopulations - a
view which has informed the dual-pathway models of basal ganglia
functional anatomy \citep{ALBIN:1989}. To date, all of our
computational models of the basal ganglia have also subscribed to this
viewpoint \citep{GURNEY:2001,HUMPHRIES:2002,GURNEY:2004}.

Specifically, we assumed that dopamine acts to gate the input to the
striatal spiny neurons, attenuating the input in D2 receptor neurons
and enhancing it in D1 receptor neurons. In our previous models, the
dopamine level was represented by two parameters: $\lambda_e$ for the D1 receptor
striatal units and $\lambda_g$ for the D2 receptor striatal units
(that is, for the selection and control pathways, respectively --- see \citet{GURNEY:2001} for details). The
gating effect of dopamine was simulated by using it as a
multiplicative factor on the weight vector {\bf w} of the afferent
connections. Thus, given the afferent input vector {\bf c}, the
activations at equilibrium $\tilde{a}_i^e$ and $\tilde{a}_i^g$ of the
$i$th striatal D1 and D2 receptor units, respectively, were given by
\begin{align}
  \tilde{a}_i^e & = \mathbf{c} \cdot \mathbf{w} \; (1 + \lambda_e) \\
  \tilde{a}_i^g & = \mathbf{c} \cdot \mathbf{w} \; (1 - \lambda_g)
\end{align}

Using this form of gating assumes that the dopaminergic synapses
formed by the substantia nigra pars compacta on spiny neurons are
physically positioned after the synapses of other afferent inputs
(particularly the glutamatergic input from cortex) and are thus able
to affect the synaptic currents generated by these afferent inputs
before they reach the soma.  However, it is highly unlikely that
sufficient dopaminergic synapses exist to meet this assumption: Recent
estimates have suggested that as little 4\% of spiny neuron dendritic
spines receive dopaminergic input, whereas 100\% of spines receive at
least one excitatory input \citep{BENNETT:2000}.  Furthermore,
dopaminergic input does not elicit a post-synaptic current and so
could not electrically gate the input from other afferents.

How then is dopaminergic input to a striatal spiny neuron able to
either attenuate or enhance the neuron's output, depending upon the
neuron's (predominant) receptor type? We leave aside for now the
question of whether the dopamine receptor sub-types do indeed have
opposite effects on spiny neurons: it will be seen below that recent
studies describing the mechanism of dopamine's action have also
provided strong evidence for this supposition. Our wish is to create a
high-level simulation (in keeping with our prior computational models)
of dopamine's affect on the output of spiny neurons via both D1 and D2
receptors and to capture how this effect changes with alterations in
dopamine level.

As detailed above, our previous model of dopamine's action focused on
the input to a model neuron. Here we begin by proposing that the
effects of dopamine are best captured by directly modeling the affect
dopamine ultimately has on the output of the neuron, thereby ignoring
the precise locus of its influence --- an approach that may be
termed {\em phenomenological} modeling \citep{HUMPHRIES:2001}. This
approach to modeling dopamine has precedent: a prior simple neural
network model emulated the effect of increased dopamine by an increase
in the slope of a model unit's output function
\citep{SERVAN-SCHREIBER:1990}. This network was thus able to replicate
the finding of increased signal detection performance in a common
psychological task (the continuous performance test) following an
increase in dopamine level.

\begin{figure}[hhh]
\begin{center}
\leavevmode
\includegraphics[width=0.5\textwidth]{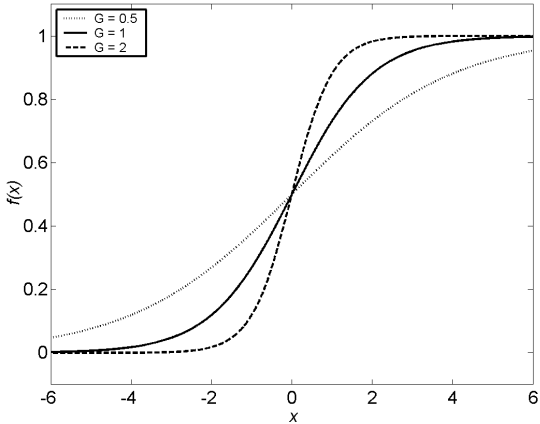}
\caption{\label{fig:sigmoid} Sigmoid functions relating neuron activation
  $x$ to normalised output $f(x)$. Increasing the slope parameter $G$ improves
  responsiveness to excitatory (positive) inputs. Note that the
  functions pivot about the fixed point $f(x) = 0.5$. }
\end{center}
\end{figure}

Servan-Schreiber et al \citep{SERVAN-SCHREIBER:1990} attributed the dopamine-mediated increase in a
neuron's responsiveness to an increase in the neuron's signal-to-noise
ratio (which would lead to improved discrimination of signal and
non-signal events and, in turn, to improved signal detection in the
test). By assuming that a neuron's output $f(x)$ could be modeled as
a sigmoid function of its input $x$
\begin{equation}
f(x) = \frac{1}{1 + e^{-(Gx+B)}}
\end{equation}
they argued that increasing the slope parameter $G$ (as illustrated in
Figure \ref{fig:sigmoid}) of this function acts to increase that
neuron's signal-to-noise ratio. Further, they showed that any
improvement in signal detection performance resulting from the
increase requires a minimum circuit of two serially linked units, an
input unit for which $G$ may be altered and an output unit which
measures signal detection performance (where signal detection was
defined as the output of the second unit crossing a threshold).

Building on this work, we took as our base assumption that increasing
dopamine will alter the slope of a neuron's output function and
extended the model of \citep{SERVAN-SCHREIBER:1990} in two directions. First,
their model took no account of the D1 and D2 receptor distinction
which crucially underpins our model of basal ganglia function.
Therefore, we wished to develop separate output models for striatal
spiny neurons with these receptor types. Second, though their
connection of psychological results with actual neural output was
impressive, we wished to explore a little deeper than the systems
level and justify the changes in output function in terms of
voltage-dependent ion channels affecting a spiny neuron's membrane
potential (and thus its output). Thus we would be able to link data
from psychology directly to a low-level description of neural
mechanisms. Further, it turned out that data on the currents affected
by dopamine facilitated our development of separate models for D1 and
D2 receptor neurons.

Having developed output models for D1 and D2 receptor neurons, we went
on to implement them in both our intrinsic \citep{GURNEY:2001} and extended \cite{HUMPHRIES:2002} models of basal ganglia function --- the extended model embeds the basal ganglia in a thalamo-cortical loop. Simulations of these functional models
demonstrated that, with appropriate parameter values, the changes made
to the model striatal neurons did not markedly affect either the model
basal ganglia's ability to act as an action selection mechanism or its
changes in output following alteration of dopamine level.

\section{Methods}
\subsection{Original neuron output model}
The simple leaky-integrator neuron used in our systems-level models
had a piece-wise linear output function, that is, a ramp which
replicated the strictly increasing portion of a sigmoid function.
Thus, the output $y$ of a single neuron given its
activation $a$ was given by
\begin{equation}
\label{equation:original_output}
y =
        \begin{cases}
                0, & \text{if } a < \epsilon;     \\
                m( a- \epsilon ),  & \text{if } \epsilon \leq a \leq 1/m + \epsilon \\
                1, & \text{if } a > 1/m + \epsilon
        \end{cases}
\end{equation}
where $m$ is the slope of the ramp, and $\epsilon$ is a threshold
term: giving it a negative value ensures a tonic output from that
neuron. In our prior models, striatal units had a threshold of
$\epsilon = 0.2$ to simulate the down-state of their biological
counterparts. Thus, the minimum activation level of 0.2 required for
striatal unit output was taken to be commensurate with the
co-ordinated cortical output required to force a spiny neuron into
its firing-ready up-state \citep{WILSON:1996}. For simplicity $m = 1$
in all of our published models to date.

\subsection{Currents of striatal neurons affected by dopamine}
The down-state of a spiny neuron is maintained by an inwardly
rectifying potassium current (KIR) which holds the neuron's membrane
potential at approximately -75 mV \citep{NISENBAUM:1995}. There is some
evidence from {\em in vitro} studies that activation of D1 receptors
by dopamine or D1-specific dopamine agonists enhances KIR
\citep{PACHECO-CANO:1996,SURMEIER:1993} thus driving the neuron's
membrane potential towards the KIR's reversal potential. In addition,
while in the down-state, D1 receptor activation attenuates a
depolarising Na$^+$ current \citep{SURMEIER:1992}. Therefore, an
increase in dopamine level would enhance a D1-type spiny neuron's
down-state and make it less responsive to excitatory input, thereby
increasing the level of co-ordinated cortical input required to cause
the up-state transition.

Once in the up-state, a spiny neuron's responsiveness to its inputs is
modulated by an L-type Ca$^{2+}$ current, and three outwardly
rectifying potassium currents \citep{BENNETT:2000}. Dopamine has a
direct effect on the L-type Ca$^{2+}$ current via both D1 and D2
receptors \citep{GREENGARD:1999}. At D1 receptors, dopamine acts to
increase the depolarising L-type Ca$^{2+}$ current, thus increasing
the neuron's responsiveness to excitatory input
\citep{HERNANDEZ-LOPEZ:1997}. Conversely, at D2 receptors, dopamine
acts to decrease the L-type Ca$^{2+}$ current, thus making the neuron
less responsive to excitatory input \citep{HERNANDEZ-LOPEZ:2000}.

The differential effects of D1 and D2 receptors on the L-type
Ca$^{2+}$ current would lead to increased and decreased GABA release,
respectively, just as reported in rat striatum \citep{O'CONNOR:1998}.
Therefore, the long-held assumption that activation of D1 receptors
excites and D2 receptors inhibits spiny neurons remains intact.
Further, the membrane-potential dependent effects of D1 receptor
activation on a striatal neuron's ion-channels accounts for the
oft-cited controversy of whether D1 receptors mediated enhanced
\citep{UMEMIYA:1997,GONON:1997} or attenuated \citep{CALABRESI:1987}
striatal output.

An intriguing behavioural study of genetically altered mice has
provided evidence that the dual effect of D1-receptor activation on
striatal output can in turn affect motor activity. Both mutant mice
which lack the KIR channel (KIR knockout) and wild-type mice
(controls) show significantly elevated locomotor activity following D1
receptor agonist injection \citep{BLEDNOV:2002}. Just such an increase
is predicted by our model of basal ganglia function: increased D1
receptor activation would lead to increased striatal output, causing
decreased SNr/GPi output, and thus increased activity via
disinhibition of motor-related structures. More pertinently, the
KIR-knockout mice had significantly elevated locomotor activity
compared to the wild-type mice following the agonist injection. The
additional increase of activity in the absence of the KIR channel is
consistent with D1's role in simultaneously hyperpolarising spiny
neurons via increased KIR while depolarising them via L-type Ca$^{2+}$
currents.

\subsection{New neuron output model}
\label{sec:new_da_model}
\begin{figure}[hhh]
\begin{center}
\leavevmode
\includegraphics[width=0.5\textwidth]{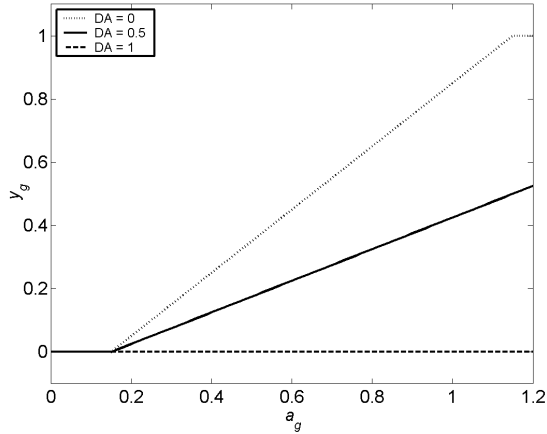}
\caption{\label{fig:D2_output_slopes} Example outputs slopes for D2
  striatal units for $\lambda_g = 0, 0.5, 1$. For these examples,
  $\epsilon = 0.15$, $m_I = 1$, and $\gamma_g = 1$.}
\end{center}
\end{figure}

Taking the D2 striatal units first, we propose that the attenuation of
the depolarising L-type Ca$^{2+}$ current can be captured by a
decrease in the slope of the output with increasing dopamine level.
Thus, the output $y_g$ of the D2 striatal units is as described by
equation (\ref{equation:original_output}), except that slope $m_g$ is
now expressed as
\begin{equation}
\label{equation:D2_slope}
m_g = m_{I} - \gamma_g \lambda_g,
\end{equation}
where $\lambda_g$ is the dopamine level in the control pathway (as
before), $m_I$ is the initial value of the slope, and $\gamma_g$ is a scaling
factor. Therefore, increasing dopamine levels correspondingly
increases the level of input required to achieve the same output, as
illustrated in Figure \ref{fig:D2_output_slopes}

\begin{figure}[h]
\begin{center}
\leavevmode
\includegraphics[width=0.5\textwidth]{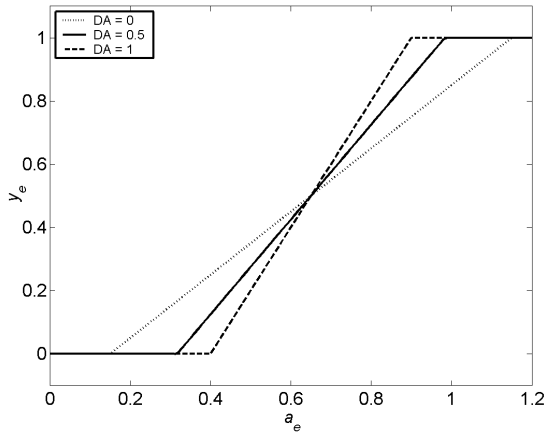}
\caption{\label{fig:D1_output_slopes} Example outputs slopes for D1
  striatal units for $\lambda_e = 0, 0.5, 1$. For these examples,
  $\epsilon = 0.15$, $m_I = 1$, $\gamma_e = 1$, and $p = 0.5$. The
  slopes pivot around the output point of 0.5 and thus increasing
  dopamine increases the amount of activation required to achieve any
  output while simultaneously increasing the level of output if the
  activation is sufficiently high. Note that $p = 0.5$ means the KIR
  and L-type Ca$^{2+}$ currents have an equal magnitude effect on the
  unit's output.}
\end{center}
\end{figure}

For the D1 striatal units two current effects must be captured, namely
the enhancements of both the hyperpolarising KIR and depolarising
L-type Ca$^{2+}$ currents following an increase in dopamine level (for
simplicity we assume that the attenuation of the Na$^+$ current is
subsumed by the enhancement of the KIR current as both changes have
the same end effect). We propose that these antagonistic affects on a
neuron's output can be captured by increasing the slope of the output
around a fixed pivot point. To do this, we need to both specify an
expression for the slope $m$ and rewrite the output expression
(\ref{equation:original_output}). First, a linear increase in slope
$m_e$ with an increase in dopamine
\begin{equation}
\label{equation:D1_slope}
m_e = m_{I} + \gamma_e \lambda_e,
\end{equation}
where, again, $\lambda_e$ is the level of dopamine in the selection
pathway, $\gamma_e$ is a scaling factor, and $m_I$ is the initial
value of the slope. Second, we add an intercept term to the second
output case which sets the pivot point, and thus also have to rewrite
its condition
\begin{equation}
\label{equation:D1_output}
y_e =
        \begin{cases}
                0, & \text{if } a < \epsilon;     \\
                m_e( a - \epsilon ) + (1 - m_e)p,  & \text{if } \epsilon \leq a \leq (1-(1-m_e)p)/m_e + \epsilon \\
                1, & \text{if } a > (1-(1-m_e)p)/m_e + \epsilon
        \end{cases}
\end{equation}
where $p$ is the desired pivot point of the output, and effectively
controls how much each of the KIR and L-type Ca$^{2+}$ channels
contribute to the overall output of the unit: setting $p > 0.5$
implies that the KIR has a greater influence; similarly, setting $p <
0.5$ implies that the L-type Ca$^{2+}$ current has a greater
influence. Thus, increasing dopamine levels both increases the level
of input required to overcome the down-state and decreases the level
of input required for the same output (if the down- to up-state
transition occurs), as illustrated in Figure
\ref{fig:D1_output_slopes}.
As can be seen from Figure \ref{fig:D1_output_slopes}, $\epsilon$ may
be more properly called the {\em initial} threshold for the D1 units,
as the activation required for any output (and thus the implied
threshold) increases with increasing dopamine.

Our proposed D1 unit output function is, with $p = 0.5$, a piece-wise
linear approximation of the dopamine model proposed by
\citep{SERVAN-SCHREIBER:1990} described above (compare Figures
\ref{fig:sigmoid} and \ref{fig:D1_output_slopes}). Based on our
simulation results, this will allow us to examine the connection they
proposed between improved signal-to-noise ratio and increased
dopamine. However, the models were motivated by different underlying
premises and thus, whereas their output function necessarily pivoted
around $f(x) = 0.5$, our approach is more flexible because it allows
for the different effects of the two currents modulated by D1 receptor
activation. In addition, we have explicitly created separate output
functions for D1 and D2 units in contrast to their general dopamine
model.

\subsection{Parameters}
As a starting point for specifying the parameters for the new output
models, we considered what the output of striatal projection neurons
would be in the absence of dopamine. Without dopamine there can be no
affect on the KIR, Na$^+$, or L-type Ca$^{2+}$ currents. Further,
there is no reason to believe that the D1 and D2 receptor neurons are
otherwise different in any way \citep{YUNG:2000}. Therefore we assumed
that for any given input the outputs of D1 and D2 striatal neurons
would be identical in the absence of dopamine. In terms of the output
models, this implies that both $\epsilon$ and $m_I$ should be
identical for the D1 and D2 striatal model units so that they have the
same output function when $\lambda_e = \lambda_g = 0$. Thus, we set
$m_I = 1$ and $\epsilon = 0.1$ for both D1 and D2 striatal units in
all the simulations.

In the absence of any contrary data, we assume that the output of D2
receptor spiny neurons does not go to zero following dopamine
saturation, and thus we must set $\gamma_g < m_I$ so that there is
model D2 unit output at maximum tonic dopamine levels ($\lambda_g =
1$).  Thus, we set $\gamma_g = 0.8$ for all the simulations reported
here. In addition, we must also assume (in the absence of contrary
data) that the magnitude of dopamine's effect is the same for both D1
and D2 receptor neurons, and thus we set $\gamma_e = 0.8$.

No data is available on which of the KIR and L-type Ca$^{2+}$ channels
have the greater effect on a striatal neuron's membrane potential
(and, therefore, its output). Thus, we will explore the effects of
varying $p$ on the selection and switching capabilities of our
intrinsic and extended basal ganglia models so that we may determine
which balance of current influence results in optimal behaviour.

Finally, all other parameters (weights, slopes, and so on) were set
according to the values given in \citep{GURNEY:2001B} for the intrinsic
model simulations, and those given in \citep{HUMPHRIES:2002} for the
extended model simulations.

\section{Results}
\label{sec:results}
As discussed above, we wished to determine how the new dopamine models
affected the ability of the basal ganglia to act as a selection and
switching mechanism. Therefore, we tested both the intrinsic and
extended models with different dopamine and pivot values using a
specified set of salience inputs. Dopamine ($\lambda_g, \lambda_e$) and
the pivot $p$ were varied over the interval [0,1] in steps of 0.1,
giving 121 pairs of dopamine and pivot values to be tested. For each
dopamine and pivot value pair, we tested salience inputs $c_1$ and
$c_2$ over the interval [0,1] in steps of 0.1, that is, for 121 pairs
of salience inputs.  Thus, both models were simulated a total of 14641
times, once for each dopamine, pivot, and salience input combination.
The input to channel 1 began at time $t$ = 1, the input to channel 2
began at $t$ = 2, and no other channels received input. This gave two
time intervals in which basal ganglia output $y^b_i$ on channel $i$ could change: $I_1 =
[1 \leq t \leq 2]$ and $I_2 = [t > 2]$.  For the intrinsic model,
salience was directly input to the striatal units; for the extended
model, salience was input to both the motor cortex and striatal units.

Selection of channel $i$ occurred when $y^b_i$ fell below a
selection threshold $\theta_s$, which we set at 0.05 in line with our
previous work. Using this definition, and given the onset times of the
salience input, the outcome of a simulation could be characterised by
one of four states:
\begin{itemize}
\item First, {\em no selection}, where $y^b_1$, $y^b_2 > \theta_s$
  for all $t$. Neither active channel becomes selected during the
  simulation.
\item Second, {\em single channel selection}, where $y^b_1 \leq
  \theta_s$ in $I_1$ and $y^b_2 > \theta_s$ in $I_2$; or $y^b_1 >
  \theta_s$ for all $t$ and $y^b_2 \leq \theta_s$ in $I_2$. A single
  channel is selected at some point in the duration of the simulation:
  either channel 1 becomes selected in the first interval or channel 2
  becomes selected in the second interval.
\item Third, {\em simultaneous channel selection}, where $y^b_1$,
  $y^b_2 \leq \theta_s$ in $I_2$. Concurrent channel selection occurs
  in the second interval (channel 1 must be selected in the first
  interval to remain selected in the second interval; but note that
  selection in the first interval does not mean automatic selection in
  the second interval).
\item Fourth, {\em channel switching}, where $y^b_1 \leq
  \theta_s$ in $I_1$ and $y^b_1 > \theta_s$ in $I_2$ and $y^b_2 \leq
  \theta_s$ in $I_2$. A clean switch between channels: channel 1 is
  selected in the first interval, then becomes de-selected as channel
  2 becomes selected in the second interval.
\end{itemize}
The outputs of a complete set of salience input pair simulations were
summarised by tallying the number of input pairs which resulted in
each of the four states. Thus, the question under study can now be
broken down into three parts: does implementing the new dopamine
mechanism result, first, in sufficient single channel selection and
switching for `normal' dopamine levels, second, in realistic changes
of the frequency of all four states with changes in dopamine levels
and, third, in these two outcomes being achieved using biologically
realistic pivot values?

\begin{sidewaysfigure}[!hp]
\begin{center}
\leavevmode
\setlength{\abovecaptionskip}{0.5cm}
\includegraphics[width=0.5\textheight]{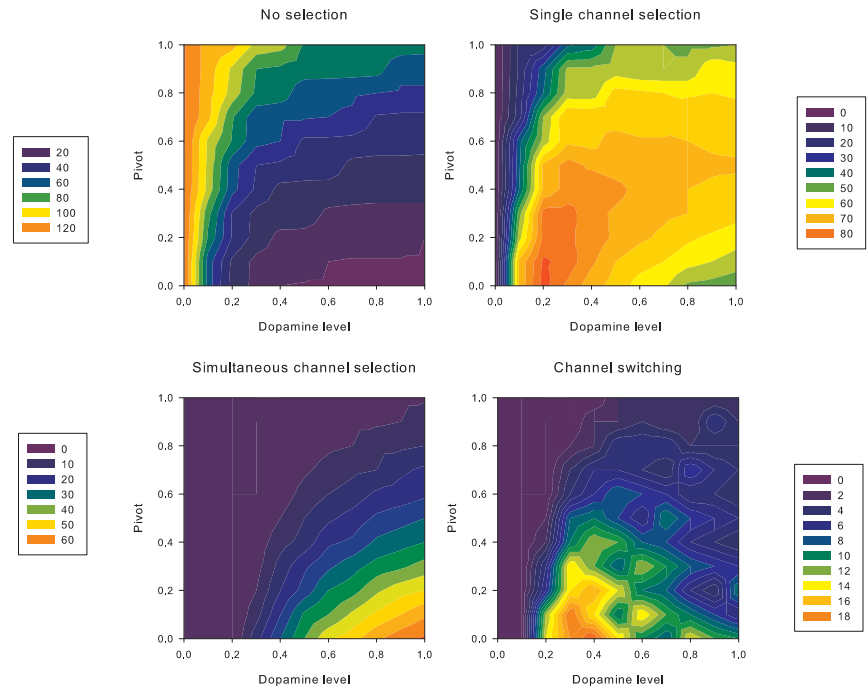}
\caption{\label{fig:intrinsic_summary} Selection and switching
  properties of the intrinsic model using the new striatal unit output
  functions.}
\end{center}
\end{sidewaysfigure}

For both the intrinsic and extended models, the answer to all of this
is yes. Taking the intrinsic model first, Figure
\ref{fig:intrinsic_summary}
shows that peak channel switching (between 14-19 input pairs)
predominantly occurred with moderate dopamine levels of between 0.3
and 0.5 and occurred at low pivot values ($p \leq 0.3$). Single
channel selection frequency within these dopamine level and pivot
boundaries was also high (63-78 input pairs), though the peak
frequency was at $\lambda_e = \lambda_g = 0.2$. Therefore, the results
from the new intrinsic model suggest that the effect of the L-type
Ca$^{2+}$ current must substantially exceed that of the KIR current
for the basal ganglia as a whole to successfully operate as an action
selection mechanism.

No selection or switching occurred with no dopamine (Figure
\ref{fig:intrinsic_summary}) as predicted by our model and consistent
with the hypothesis that the difficulty of initiating voluntary
movement for parkinsonian patients results from an inability to remove
the tonic inhibition by basal ganglia output of motor structures in the thalamus
and brainstem. Increasing dopamine to high levels, as might occur
following amphetamine stimulation, resulted in a substantial increase
in simultaneous channel selection which is indicative of multiply
selected or alternated actions. Thus, the model's output following
changing dopamine levels does correlate with behavioural problems
observed in extreme dopamine conditions. However, again the most
accurate outputs were only observed for low pivot values, with
simultaneous channel selection only substantially increasing at $p \leq
0.4$.

\begin{sidewaysfigure}[!hp]
\begin{center}
\leavevmode
\setlength{\abovecaptionskip}{0.5cm}
\includegraphics[width=0.5\textheight]{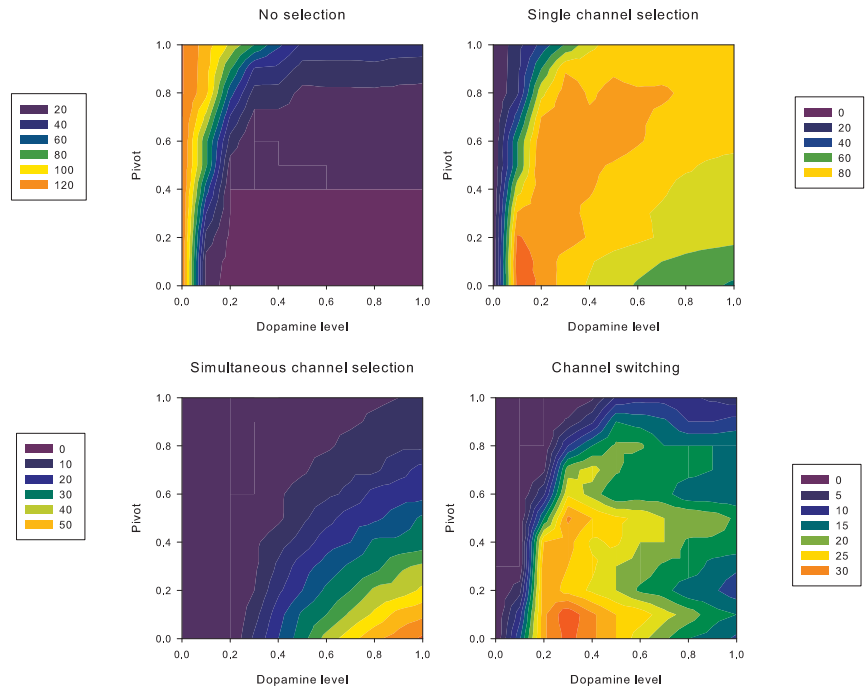}
\caption{\label{fig:extended_summary} Selection and switching
  properties of the extended model using the new striatal unit output
  functions.}
\end{center}
\end{sidewaysfigure}

The extended model results, shown in Figure
\ref{fig:extended_summary}, are surprisingly similar to those of the
intrinsic model in that the peak channel switching (25-32 input pairs)
occurred with dopamine levels between 0.2 and 0.5, high frequency
single channel selection occurred at these dopamine levels, extreme
low and high dopamine conditions resulted, respectively, in no
selection and high frequency simultaneous channel selection, and all
of these results required low pivot values ($p \leq 0.5$). The major
difference was that channel switching occurred more often for every
dopamine and pivot value pair for the extended model, and that the
frequency of channel switching in the extended model exceeded the peak
frequency of channel switching in the intrinsic model for 33 parameter
pairs, within the pivot range (0.0-0.7) and dopamine values (0.2-0.8).
However, beyond $p = 0.5$ there were only two dopamine-and-pivot pairs
which showed greater frequency of channel switching than the peak
found in the intrinsic model. Thus, in the extended model the basal
ganglia's operation as an action selection mechanism was more robust
to changes in dopamine and pivot levels.  Therefore, the extended
model allows for a more flexible hypothesis regarding the balance of
the L-type Ca$^{2+}$ and KIR currents in that anything from equal
current contributions to exclusively L-type Ca$^{2+}$ current
contributions results in sensible selection behaviour.

Having examined the results from the new versions of the intrinsic and
extended models, we now need to compare these results with those of
the original models to draw firmer conclusions on how the new striatal
unit output models have affected the selection and switching
properties. To do this we must first identify the parameter set
$\{ \lambda_e, \lambda_g, p \}$ which resulted in optimal selection and
switching performance in the new versions, so that we are able to
compare individual simulation sets (that is, a complete set of
salience input pairs) across the models. A simple measure of the
success of a particular parameter set is the ratio $R$ between the number
of input pairs which resulted in successful selection or switching and
the number of input pairs which did not (in the simulation set which
used those parameters). Expressed formally, this is simply:
\begin{equation}
\label{equation:optimisation}
R = \frac{F_b + F_d}{F_a + F_c},
\end{equation}
where $F_a$, $F_b$, $F_c$, and $F_d$ are the frequencies of no selection,
selection, simultaneous channel selection, and channel switching,
respectively for a simulation set. Thus, the optimal parameter set
is that for which $R$ is maximised, constrained by the boundary
conditions $0 < p < 1$, because we know that both the KIR and L-type
Ca$^{2+}$ channels affect the neuron to some degree, and $0 < \lambda_e,
\lambda_g < 1$ because we know that dopamine is not absent from or
saturated in striatum under normal conditions.

\begin{figure}[!htp]
\begin{center}
  \leavevmode
  \includegraphics[width=0.5\textwidth]{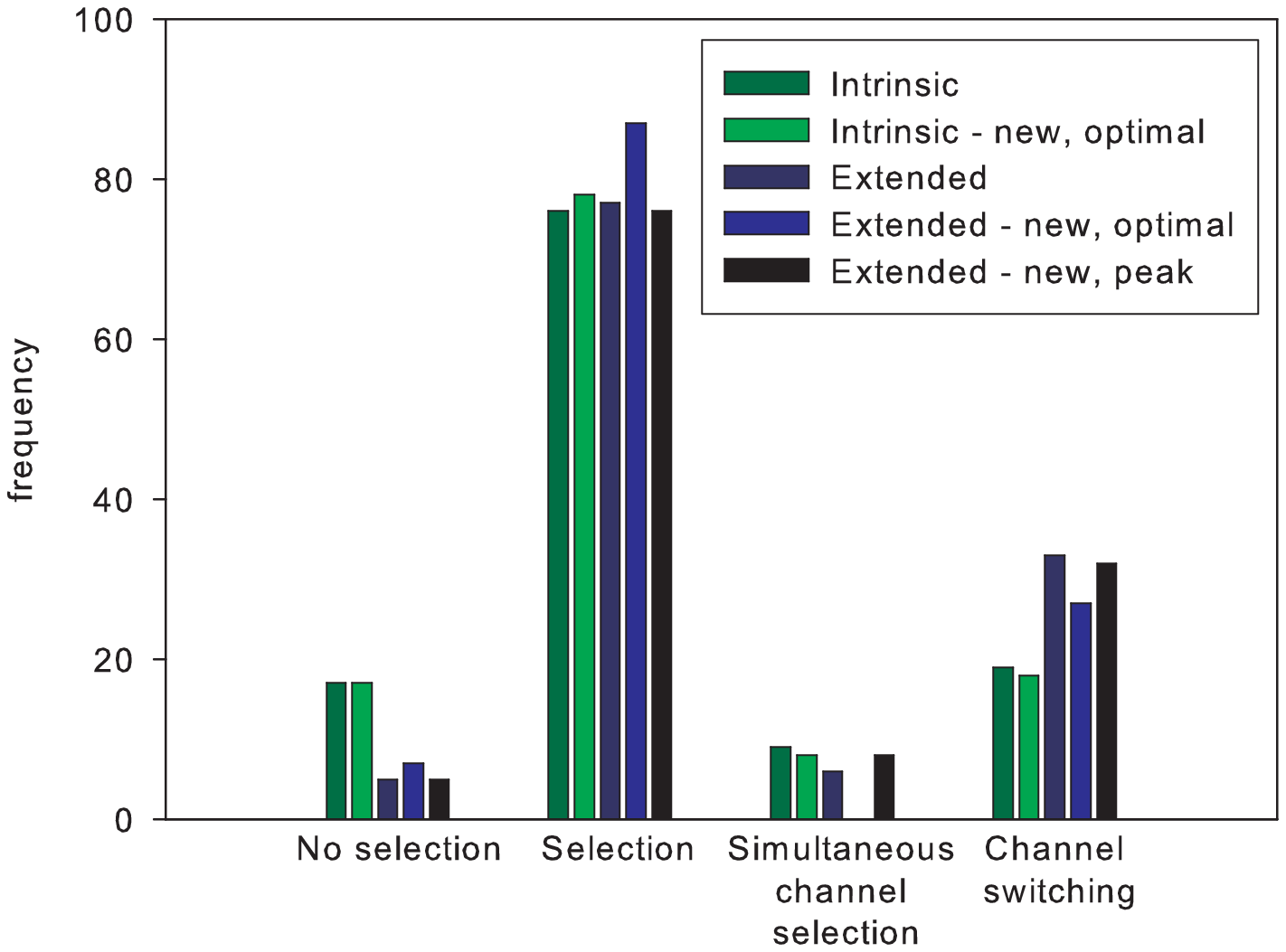}
\caption{\label{fig:comparison_hist} Histogram of the frequency of the
  four basal ganglia output states resulting from complete simulation sets run on
  the original and new (optimal) extended and intrinsic models and
  from the new extended model whose parameter set results in the
  greatest switching frequency. The frequencies of the output states
  are almost identical for both versions of the intrinsic model, and
  for the extended and new peak-switching extended models.
  Simultaneous channel selection does not occur at all for the new
  (optimal) extended model, indicating that this model has the
  property of ``hard switching''.}
\end{center}
\end{figure}

For the new versions of the models, maximal $R$ occurs at $p = 0.1,
\lambda_e = \lambda_g = 0.3$ for the intrinsic model and $p = 0.1,
\lambda_e = \lambda_g = 0.2$ for the extended model. Figure
\ref{fig:comparison_hist} shows the comparative frequencies of the
four EP output states of the simulation sets corresponding to these
parameters and of the original intrinsic and extended models.
It can be seen from the histogram that the frequencies of the output
states of the original and (optimal) new versions of the intrinsic
model are almost identical. Thus, the use of the new striatal unit
output functions has not altered the selection and switching
properties of the intrinsic model.

The performance of the (optimal) new version of the extended model
does differ from the original version. Most striking is that the new
version showed no simultaneous channel selection at all, indicating
that this particular implementation of the model only ever allows a
maximum of one channel to be selected at any given moment - a property
which may be termed ``hard switching''. However, if we consider the
parameter set which gives the maximum switching frequency ($p = 0.1,
\lambda_e = \lambda_g = 0.3$, the same values as the optimal intrinsic
model) then, as Figure \ref{fig:comparison_hist} shows, the
frequencies of the output states of the original and (optimal) new
versions of the extended model are again almost identical. Thus, we
can conclude that the selection and switching properties of the
original extended model {\em can} be maintained when using the new
striatal unit output functions but there is also a different, optimal
parameter set which results in hard-switching.

Optimal selection and switching performance was achieved using the
lowest possible pivot value ($p = 0.1$) for both the intrinsic and
extended models, which implies that, for optimal performance, the
influence of the L-type Ca$^{2+}$ current on the unit's output must
far exceed that of the KIR current. However, as we noted when
considering the results of the extended model, near-optimal
performance, with high frequencies of switching, can be achieved with
a wide range of pivot values.

\subsection{Signal detection performance of the new output models}
We briefly noted above that the close approximation of our D1 unit
output model to Servan-Schreiber et al's \citep{SERVAN-SCHREIBER:1990} general dopamine model
would allow us to examine their claim that increased dopamine improves
the signal-to-noise ratio at both cellular and behavioural levels by
increasing the slope of a neuron's output function. More specifically,
they demonstrated that by increasing the slope of their (sigmoid)
output function, their network model of the continuous performance
test was better able to discriminate between signal and non-signal
events by virtue of decreasing the number of missed signal events.
In addition, they noted that a minimum network of two serial units was
required to observe the improvement in signal detection performance.

\begin{figure}[!hp]
\begin{center}
\leavevmode
\includegraphics[width=0.5\textwidth]{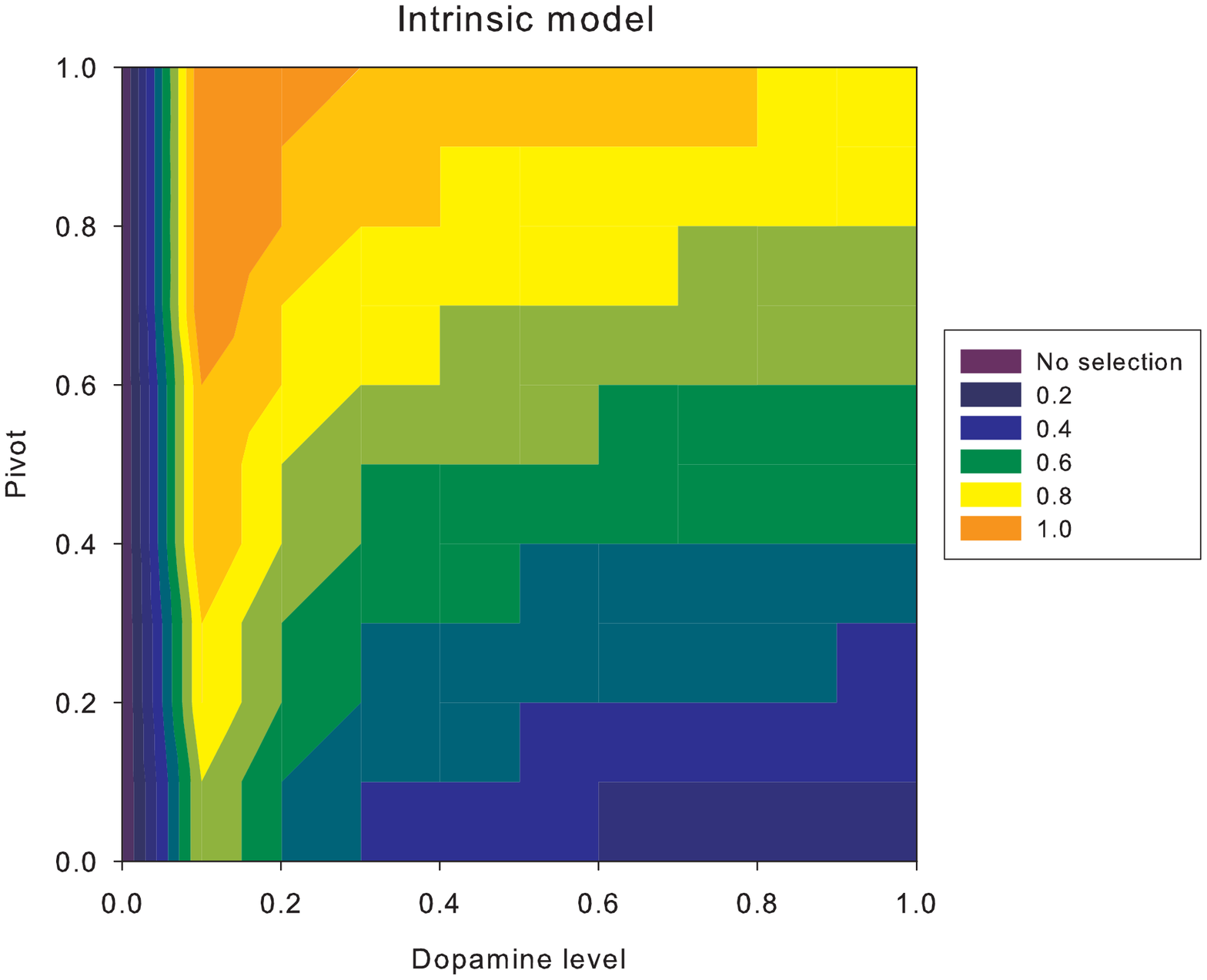}
\includegraphics[width=0.5\textwidth]{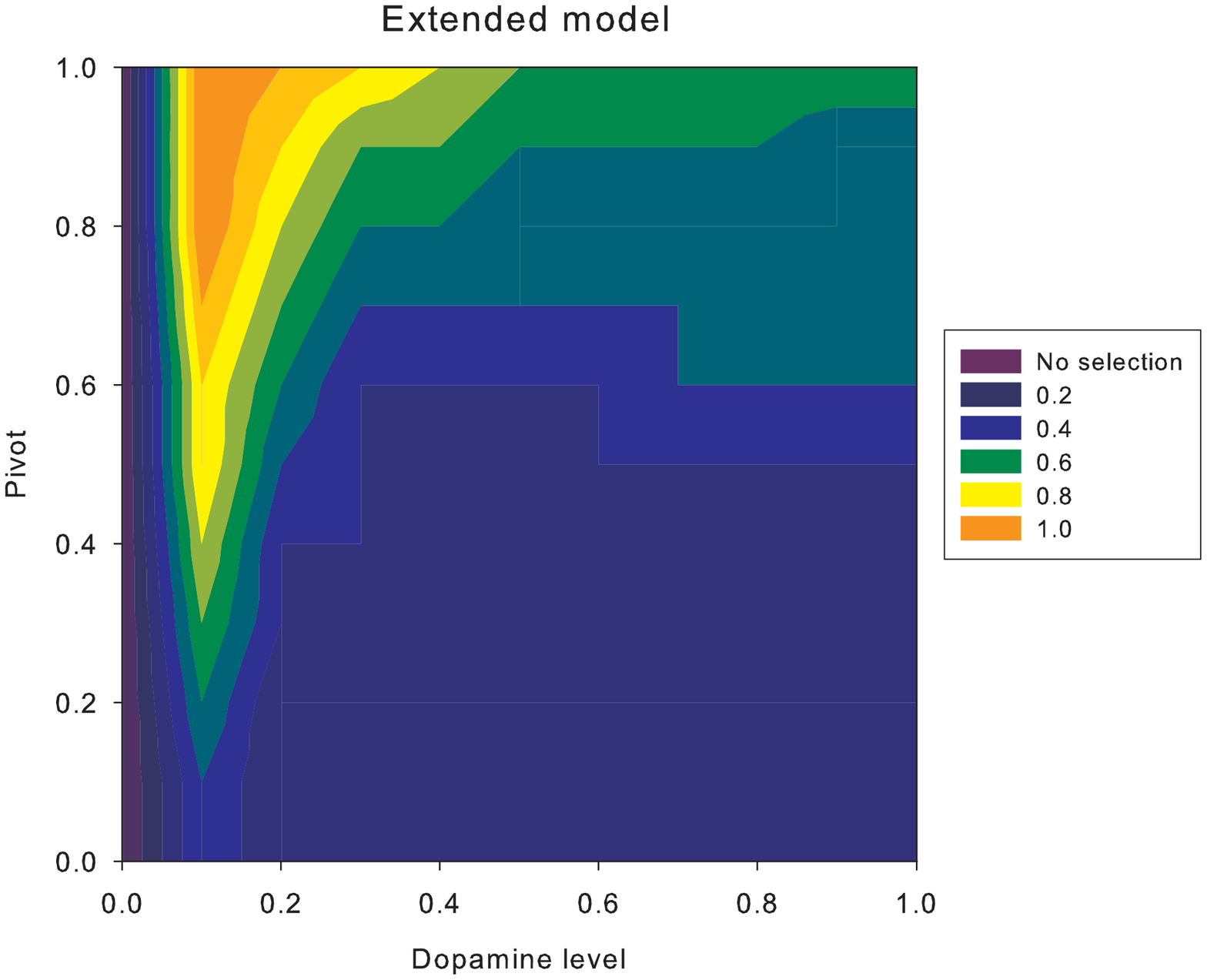}
\caption{\label{fig:min_salience} Minimum salience level required for
  selection of a single channel in the intrinsic (top) and extended
  (bottom) models. }
\end{center}
\end{figure}

Translating these findings into our action selection framework, we
note direct correspondences with our models: that our signal input is
the salience level of the action, that signal detection is indicated
by a selected channel, and that our network fulfills the requirement
of having the minimum network: an input unit for which the slope is
altered (here the striatal D1 and D2 units) and an output unit to
observe the signal detection (here the basal ganglia output nuclei). Thus, we predict
that increasing the level of dopamine (and thereby altering the slopes
of the striatal unit output functions) would lead to increased
detection of salient stimuli. This is exactly what we have found: as
illustrated in Figure \ref{fig:min_salience}
the minimum input value required for single channel selection
decreases with increasing dopamine for every pivot value in both the
intrinsic and extended models. Thus, increasing dopamine improves both
models' ability to detect signal (salient) events.

It is worth noting that the improved signal detection performance
following increased dopamine occurred even though our models utilised
two types of striatal units with opposite effects on slope, in
contrast to Servan-Schreiber et al's model \citep{SERVAN-SCHREIBER:1990} which only contained
a model roughly equivalent to our D1 model unit. Moreover, the slope
of their output function pivoted at 0.5, which in the context of the
models presented here would imply an equal influence of the KIR and
L-type Ca$^{2+}$ currents on the D1 neuron's membrane potential. We
have shown that the improvement in signal detection can occur at any
pivot point and is, therefore, independent of the exact combination of
influences of these two currents.  Thus, by using a more accurate
model of dopamine's effects we have been able to provide strong
evidence that increasing dopamine will improve signal detection
performance.

\section{Discussion}
We have demonstrated that replacing a gating model with a slope model
of dopamine's effect on striatal unit output has not significantly
altered the ability of the simulated basal ganglia to produce outputs
consistent with the selection of and switching between salient
actions. In addition, the results from the new version of the extended
model have suggested that there is an optimal basal ganglia state in
which simultaneous channel (and, therefore, action) selection is
eliminated, thus only allowing ``hard switching'' between actions.

The dopamine and D1 output function pivot values for which selection
and switching was either replicated (from the original models) or
optimal in the new models imply low-to-moderate tonic dopamine and a
greater influence of the L-type Ca$^{2+}$ current on the D1 neuron's
membrane potential than the KIR current. Thus, if the basal ganglia is
indeed a channel-based switching mechanism in the vertebrate action
selection system then we would expect to observe tonic dopamine levels
significantly below saturation and greater magnitude effects of the
L-type Ca$^{2+}$ than the KIR current in normally healthy, behaving,
animals.

Altering D1 output function pivot values did not effect the general
trend of decreasing minimum salience input required for selection with
increasing dopamine level. If this trend is taken to be indicative of
increased signal detection, then we have provided further evidence
that increasing dopamine does increase the signal-to-noise ratio of
a neuron's output by altering the neuron's output function slope, as
proposed by \citep{SERVAN-SCHREIBER:1990}.

\subsection{Assumptions of the slope model}
A critical assumption we have made is that the dopamine-modulated
ion-channels affecting D1 and D2 receptor neurons give graded response
to increases in dopamine, that is, there isn't a critical level of
dopamine at which the channels jump from one state to another, but
that their effect continuously varies. An {\em in vivo} combined
voltammetric analysis and single-unit recording study has demonstrated
that the strength of both the excitatory and inhibitory responses of
striatal medium spiny cells to dopamine release are dependent on the
concentration of dopamine (in effect, the tonic level of dopamine)
\citep{WILLIAMS:1990}. Thus, we may reasonably expect that as neural
output is a continuous function of dopamine level then the currents
underlying the generation of this output must also continuously vary
in strength.

Other assumptions of the models are open to further testing: we
assumed a linear increase in D1 neuron output slope and,
correspondingly, linear decrease in D2 neuron output slope with
increased dopamine levels purely because this is the simplest possible
relationship between slope and dopamine levels. It is certainly
possible that either or both of the neuron types have alternate
dopamine-slope relationships, such as exponential changes in output
slope with increasing dopamine. Further, it is possible that the
decrease in output slope of D2 neurons via the attenuation of the
L-type Ca$^{2+}$ current is not of equal magnitude to the increase in
output slope of D1 neurons via enhancement of the same current, as we
have modeled here. Thus, further modeling studies are required to
explore the effects of altering the slope's dopamine-response function
and altering the magnitude of dopamine's effect. (Note also that the
linear increase in output slope results in non-linear increase of
(implied) output threshold and non-linear decrease of output
saturation level for D1 units.)

In addition, the exact slope-dopamine level relationship may be
further clarified by {\em in vitro} or {\em in vivo} studies. For the
D2 neurons, the following is suggested: in slice or culture,
voltage-clamped to the up-state (so that L-type Ca$^{2+}$ channel is
active) with relatively strong current injection to ensure tonic
output of striatal neurons, introduce D2 agonist to extra-cellular
bath, and measure (hopefully) decreased output rate of striatal cells
with increasing D2 agonist levels. Due to the dual currents active for
D1 receptor neurons, a much more complex study would be required,
presumably utilising channel-specific toxins to (mostly) eliminate
each ion-channel in turn.

There is also the assumption, underlying all of our basal ganglia
models to date, of two populations of striatal neurons, divided
between D1- and D2-receptor neuron types and their respective
projection targets.  Numerous studies have reported that D1 and D2
receptors co-localise on anywhere between 35\% and 100\% of striatal
medium spiny neurons
\citep{SURMEIER:1992,SURMEIER:1996,HARSING:1997,AIZMAN:2000}. However,
the more recent studies (see, for example, Aizman et al., 2000) have
suggested that most researchers (for example, Gerfen et al., 1990)
\nocite{GERFEN:1990} do not find co-localised receptors because D2
receptor concentration is far smaller than D1 in EP/SNr projecting
neurons, and D1 concentration far smaller than D2 in GP(e) projecting
neurons. Thus, even though they may be co-localised, there is the
distinct possibility that their effect (when in minority) is
functionally insignificant. However, it would still be prudent to
explore the effects of receptor co-localisation in a basal ganglia
model, using the new striatal output functions.

\subsection{Conclusion}
To fully validate the striatal unit output functions presented here,
these results require replication at the biophysical level, at which
each ion-channel is uniquely described and their spatial effects can
be modeled. However, the proposed output functions serve as an
interim (and computationally efficient) step with which we have
further demonstrated the possibility of capturing complex current
dynamics in high-level models and have been able to relate
ion-channel effects, neural output, and psychological test data.

\bibliography{da_bib}

\begin{thebibliography}{10}

\bibitem{POIRIER:1965}
L.~J. Poirier and T.~L. Sourkes,
\newblock Brain {\bf 88}, 181 (1965).

\bibitem{ALBIN:1989}
R.~L. Albin, A.~B. Young, and J.~B. Penney,
\newblock Trends Neurosci {\bf 12}, 366 (1989).

\bibitem{GURNEY:2001}
K.~Gurney, T.~J. Prescott, and P.~Redgrave,
\newblock Biol Cybern {\bf 85}, 401 (2001).

\bibitem{HUMPHRIES:2002}
M.~D. Humphries and K.~N. Gurney,
\newblock Network {\bf 13}, 131 (2002).

\bibitem{GURNEY:2004}
K.~N. Gurney, M.~Humphries, R.~Wood, T.~J. Prescott, and P.~Redgrave,
\newblock Network {\bf 15}, 263 (2004).

\bibitem{BENNETT:2000}
B.~D. Bennett and C.~J. Wilson,
\newblock Synaptology and physiology of neostriatal neurones,
\newblock in {\em Brain dynamics and the striatal complex}, edited by R.~Miller
  and J.~Wickens, Harwood Academic Publishers, London, 2000.

\bibitem{HUMPHRIES:2001}
M.~D. Humphries and K.~N. Gurney,
\newblock Neural Netw {\bf 14}, 845 (2001).

\bibitem{SERVAN-SCHREIBER:1990}
D.~Servan-Schreiber, H.~Printz, and J.~D. Cohen,
\newblock Science {\bf 249}, 892 (1990).

\bibitem{WILSON:1996}
C.~J. Wilson and Y.~Kawaguchi,
\newblock J Neurosci {\bf 16}, 2397 (1996).

\bibitem{NISENBAUM:1995}
E.~S. Nisenbaum and C.~J. Wilson,
\newblock J Neurosci {\bf 15}, 4449 (1995).

\bibitem{PACHECO-CANO:1996}
M.~T. Pacheco-Cano, J.~Bargas, S.~Hernandez-Lopez, D.~Tapia, and E.~Galarraga,
\newblock Exp Brain Res {\bf 110}, 205 (1996).

\bibitem{SURMEIER:1993}
D.~J. Surmeier and S.~T. Kitai,
\newblock Prog Brain Res {\bf 99}, 309 (1993).

\bibitem{SURMEIER:1992}
D.~J. Surmeier et~al.,
\newblock Proc Natl Acad Sci U S A {\bf 89}, 10178 (1992).

\bibitem{GREENGARD:1999}
P.~Greengard, P.~B. Allen, and A.~C. Nairn,
\newblock Neuron {\bf 23}, 435 (1999).

\bibitem{HERNANDEZ-LOPEZ:1997}
S.~Hernandez-Lopez, J.~Bargas, D.~J. Surmeier, A.~Reyes, and E.~Galarraga,
\newblock J Neurosci {\bf 17}, 3334 (1997).

\bibitem{HERNANDEZ-LOPEZ:2000}
S.~Hernandez-Lopez et~al.,
\newblock J Neurosci {\bf 20}, 8987 (2000).

\bibitem{O'CONNOR:1998}
W.~T. O'Connor,
\newblock Nucl Med Biol {\bf 25}, 743 (1998).

\bibitem{UMEMIYA:1997}
M.~Umemiya and L.~A. Raymond,
\newblock J Neurophysiol {\bf 78}, 1248 (1997).

\bibitem{GONON:1997}
F.~Gonon,
\newblock J Neurosci {\bf 17}, 5972 (1997).

\bibitem{CALABRESI:1987}
P.~Calabresi, N.~Mercuri, P.~Stanzione, A.~Stefani, and G.~Bernardi,
\newblock Neurocience {\bf 20}, 757 (1987).

\bibitem{BLEDNOV:2002}
Y.~A. Blednov et~al.,
\newblock Psychopharmacology {\bf 159}, 370 (2002).

\bibitem{YUNG:2000}
K.~K. Yung and J.~P. Bolam,
\newblock Synapse {\bf 38}, 413 (2000).

\bibitem{GURNEY:2001B}
K.~Gurney, T.~Prescott, and P.~Redgrave,
\newblock Biol Cybern {\bf 85}, 411 (2001).

\bibitem{WILLIAMS:1990}
G.~V. Williams and J.~Millar,
\newblock Neuroscience {\bf 39}, 1 (1990).

\bibitem{SURMEIER:1996}
D.~J. Surmeier, W.~J. Song, and Z.~Yan,
\newblock J Neurosci {\bf 16}, 6579 (1996).

\bibitem{HARSING:1997}
J.~Harsing, L.~G. and M.~J. Zigmond,
\newblock Neuroscience {\bf 77}, 419 (1997).

\bibitem{AIZMAN:2000}
O.~Aizman et~al.,
\newblock Nature Neurosci {\bf 3}, 226 (2000).

\bibitem{GERFEN:1990}
C.~Gerfen et~al.,
\newblock Science {\bf 250}, 1429 (1990).

\end{thebibliography}
\end{document}